# Field-Programmable Mobile Magneto-Photonic Metaparticles for Active Light Manipulation and Steering

**Authors**

Seung Yeol Lee[1], Yujie Luo[1], Ognjen Ilic[1*]

**Affiliations**

[1] Department of Mechanical Engineering, University of Minnesota, Minneapolis, 55455 USA

**Abstract**

Controlling the flow of light within complex and dynamic environments is essential for a wide range of applications, from deep-tissue imaging and optogenetics to precision phototherapy. Typically, such light flows are controlled using external optical systems requiring line-of-sight access or by embedded nanoparticle scatterers with limited directional control, underscoring the need for mobile photonic agents capable of actively delivering and steering light within complex media. Here, we present magneto-photonic metaparticles: mobile, magnetically actuated microstructures that integrate a magnetic core with a nanoimprinted photonic surface. This hybrid design merges the reconfigurability of photonic metasurfaces with the mobility of magnetic actuation, enabling programmable translation, rotation, and real-time beam steering in aqueous media. In a concept-proof demonstration, we realize polymeric metaparticles with embedded magnetic core and nanoimprinted surface that exhibit controlled locomotion and active, magnetically programmable beam steering. Our design approach further points to more sophisticated metaparticle designs with high-efficiency, polarization-insensitive light steering, compatible with a scalable, single-step nanoimprint process. The mobile magneto-photonic metaparticle platform combines metasurface-level optical control with magnetic mobility, offering a versatile and scalable approach for active photonic control in complex environments.

## 1. Introduction

The ability to deliver light of desired properties (such as wavelength, intensity, and polarization) to precise locations in a complex medium is critical for many applications across biology and biomedicine, from neural stimulation[1–3] and imaging[4,5] to drug delivery[6] and thermotherapy.[7]. In biological and aqueous environments, this task becomes especially challenging because targets are rarely stationary: cells drift with fluid flow,[8,9] and tissues or organs exhibit micromotion.[10,11] Conventional optical systems and probes,[12] which rely on external beam steering, are poorly suited for such dynamic conditions: external illumination systems are sensitive to alignment as targets move or deform, and they require a line-of-sight access to be effective. These limitations underscore the need for embedded light-delivery agents: miniaturized, mobile platforms capable of controlling the flow of light within the environment of interest. Such in-situ control would enable real-time beam steering, allowing precise delivery and collection of light for applications ranging from optical stimulation[13] to imaging[14] and diagnostics. Achieving this, however, demands an active photonic system that combines compactness, responsiveness, and optical response, all within microscale dimensions.

One approach toward miniaturized light delivery involves using nanoparticles as mobile optical agents. Metal and dielectric nanoparticles[4,6,15–17] can scatter light within aqueous environments, and their optical response can, in principle, be harnessed for imaging or therapeutic purposes. Isotropic particles,[18] such as spherical nanoparticles, offer limited means of controlling the directional control of scattered light due to their symmetric profile. Introducing anisotropy through nanorods,[19] nanodiscs,[20] or Janus structures,[21,22] can provide partial control over polarization or directional scattering; however, the orientation of these particles in a fluidic environment is easily randomized by Brownian motion and hydrodynamic forces, making stable directional control challenging. Furthermore, the subwavelength scale of such particles fundamentally limits the optical response they can support, making it impossible to implement advanced photonic functionality, such as beam steering, in such a small form.

On the other hand, the level of photonic functionality unattainable in nanoparticles has been realized in metasurfaces—engineered structures that have attracted significant recent attention due to their unprecedented ability to shape light flows. These subwavelength-engineered surfaces[23–25] have transformed optics and photonics, enabling precise manipulation of phase, amplitude, and polarization for applications in optical communication,[26–28] imaging,[29,30] and optical sensing.[31,32] However, despite their versatility, metasurfaces are typically realized as immobile and substrate-bound devices. Even when tuned electrically,[33] thermally,[34,35] or mechanically,[36,37] they remain immobile and static, lacking the ability to operate freely within three-dimensional or aqueous environments. This immobility fundamentally restricts their use in configurations demanding in-situ adaptability or mobility.

Here, we present a new approach that combines the advanced photonic control of beam steering with the mobility of particle-based systems (Figure 1a). We introduce magneto-photonic meta-particles: embeddable, magnetically actuated microstructures that integrate a magnetic core with a nanoimprinted photonic surface. This configuration enables precise motion and orientation control through external magnetic fields while maintaining full photonic functionality for targeted light delivery. By transitioning from an immobile, macroscale and substrate-bound metasurface to a mobile, particle-based architecture, this approach merges the versatility of metasurface optics with the spatial freedom (and non-line-of-sight control) of magnetic actuation. The result is a scalable, reconfigurable photonic platform capable of real-time beam steering in dynamic environments, as conceptually illustrated in Figure 1. This multilayer microstructure integrates the desired photonic and magnetic functionalities into a mobile, embeddable platform that is fabricable at scale and offers a versatile, highly tunable photonic design space.

## 2. Results
### 2-1 Design of magneto-photonic metaparticles

We realize this design experimentally by fabricating magneto-photonic meta-particles that embody the proposed integration of magnetic actuation and photonic functionality. Each magneto-photonic meta-particle consists of a polymer base, a magnetic core, a nanoimprinted polymer scaffold, and a top layer (Figure 1b). This multi-layered architecture allows the magnetic core to govern the locomotion and orientation, while the nanoimprinted surface and the top layer create and enhance photonic functionality, respectively (Figure 1c). The magnetic core is shaped as an isosceles triangle to break radial symmetry and introduce shape anisotropy, which promotes alignment during rotation and allows the rotational motion to be easily visualized. Figure 1d shows a scalable fabrication process that assembles magnetic and photonic layers. Here, SU-8, a negative photoresist, serves as a polymer scaffold, enabling the formation of the particle body through photolithography and the scalable production of photonic surfaces via nanoimprinting lithography at the same time. Arrays of SU-8 bases are first fabricated on a silicon wafer. The magnetic core is then integrated through three sequential steps: 1) spin-coating and developing a positive photoresist on top of the SU-8 base, 2) directional deposition of a ferromagnetic material using an e-beam evaporator, and 3) lift-off of the photoresist. Because the patterning occurs within the recessed region of the polymer base, the positive photoresist is made sufficiently thick to cover the base and is developed in an inverted pattern, defining the magnetic core geometry. Subsequent metal deposition and lift-off leave the magnetic core precisely positioned on top of the SU-8 base. The top SU-8 layer is then formed by spin-coating and developing, followed by nanoimprinting photonic patterns and a deposition of a dielectric or metallic coating. Finally, magneto-photonic microstructures are detached from the substrate and collected.

## 2-2 Controlled locomotion and rotational motion of magnetic metaparticles

To demonstrate the magnetic actuation and locomotion capabilities of the magneto-photonic microdiscs, we characterized their translational motion under an external magnetic field. A neodymium magnet was used to actuate the microdiscs immersed in DI water. The position of each microdisc was tracked over time to determine the translation speed ($v$) as a function of the distance to the magnet ($d_{d-m}$). The measured velocities for $d_{d-m}$ = 2, 3, 5, 7, and 10 cm were 162.4, 25.3, 4.1, 1.0, and 0 μm/s, respectively (Figure 2a). At $d_{d-m}$ = 10 cm, the microdisc exhibited negligible displacement over 20 s, making its velocity difficult to measure. The log-log plot of velocity vs distance revealed a slope of -3.99, consistent with the $v \propto d_{d-m}^{-4}$ scaling, which can be predicted from the far-field magnetic dipole gradient ($dB/d(d_{d-m}) \propto d_{d-m}^{-4}$) and viscous drag balance for a saturated ferromagnetic inclusion in low-Reynolds-number flow (Figure 2b). To demonstrate programmable control, the microdisc structures were actuated through a sequence of three translational motions: 1) upward, 2) rightward, and 3) upward again, by sequentially switching the magnet's position (Figure 2c and movie S1). These results confirm that the magneto-photonic microdiscs exhibit controllable, programmable locomotion, and enable obstacle avoidance and targeted movement under simple magnetic field manipulation.

Next, we show that the magneto-photonic microdiscs can be both translated and rotated on demand. We implement this functionality through two sequential magnetic motions: 1) translating the metaparticle to the desired location (Figure 2c) and 2) rotating it to achieve the target beam steering direction (Figure 3a,b, and movie S2). For our particle design, distance $d_{d-m}$ between the magnet and the particle can separate motion regimes. In regime I ($d_{d-m}$ < 10 cm), the applied field is sufficiently strong to support locomotion (i.e., the first magnetic motion), translating the particle in the desired direction. In regime II (10 cm < $d_{d-m}$ < 20 cm), the applied field can only support rotation: this regime is suitable for beam steering, or second magnetic motion, where the rotation is decoupled from translation (Figure 3c). The magnetic metaparticle is oriented toward/away based on the applied field direction due to the built-in magnetic shape anisotropy (Figure 2c and Figure S1). Finally, in regime III ($d_{d-m}$ > 20 cm), the field is too weak to effectively orient the particle (i.e., the rotational rate decreases below 0.1 Hz). Together, these field-dependent regimes establish a straightforward mechanism to selectively trigger translation or rotation, providing precise and decoupled control necessary for active beam steering.

## 2-3 Experimental demonstration of active beam steering capability

To integrate photonic functionality into the metaparticle, we employ nanoimprinting, a scalable fabrication approach that leverages the deformable nature of SU-8 to enable high-throughput replication of nanoscale features. To improve yield and minimize detachment of the SU-8 film, we imprint the mold onto the metaparticle surface where the SU-8 film has been fully cross-linked. As a fabrication proof-of-concept, we imprint a grating (pitch 833 nm, see Methods), capable of generating a strong -1$^{st}$ order diffraction at 633 nm with steering angle at 49.4°. The nanoimprint process is followed by a followed by 50 nm Ag deposition (Figure S2) and detachment of metaparticles in the array (Figure 4a). Once the photonic pattern is imprinted, the magnetic and photonic orientation are synchronized, such that in-plane rotation of the metaparticle under external magnetic field determines the direction of beam steering. We note that, even if the metaparticle is accidentally oriented with its photonic surface facing downward, an out-of-plane magnetic rotation can flip it (Figure S3), restoring the active surface to the upward orientation (movie S3).

Having fabricated the composite magneto-photonic particle, we demonstrate its beam-steering capability by tracking the trajectories of light during controlled magnetic rotation. Immersed metaparticles were placed beneath a hollow, translucent hemisphere with an aperture at the top (Figure 4b). A neodymium magnet, mounted on a linearized stepper motor via an optical dovetail rail, was positioned so that its linear motion was converted into circular motion around the sample. A red laser beam (633 nm) was directed through the aperture onto the sample plane, and the directed beam spot was projected onto the inner surface of the hemisphere. For in-plane rotation, the magnet was oriented toward the metaparticle while moving circularly, producing a visible trajectory of the steered beam along the azimuthal direction (Figure 4b). The magnet was programmed to execute a sequence of motions: 1) 35° clockwise rotation, 2) 2 s pause, 3) 70° counterclockwise rotation, 4) 2 s pause, and 5) 35° clockwise rotation. The diffracted beam spot followed these programmed rotations immediately and without delay (Figure 4c,d, and movie S4), confirming that the photonic orientation precisely tracks the magnetic orientation of the particle.

Observing that the magnetic and photonic orientations are synchronized, we proceed to demonstrate active beam steering of the magneto-photonic metaparticle. The experimental setup consists of three light-sensor boards, mounted on the three walls of the container, with the light sensors facing inward. On each board, opposite of the light sensor, an outward facing LED indicates whether the corresponding sensor is triggered (Figure 4e). Red laser light, directed through a small aperture in the container top, is steered by the metaparticle with its azimuthal steering angle controlled by the circular motion of an external magnet. The light-sensor boards are programmed to activate the corresponding green LED whenever the associated sensor is triggered, serving as an indicator that the steered beam has hit the target sensor (see Methods). Using the same programmed sequence of circular magnet motions as in Figure 3d, this setup produces sequential LED activations (Figure 4f and movie S5). These results validate real-time, magnetically driven beam steering and demonstrate the rapid and synchronized response of the magneto-photonic metaparticle to external magnetic fields.

**2-4 Nanoimprinting-compatible optimized metagratings design**
Building on the proof-of-concept demonstrations, we next extend our approach toward a nanophotonic design that fully exploits the fabrication scalability offered by nanoimprint lithography *while* remaining compatible with fabrication constraints. Achieving high-performance metaparticle nanophotonic designs compatible with nanoimprint processes is a key challenge. Although nanoimprint lithography enables large-area fabrication of photonic structures, its reliance on polymer materials with low to moderate refractive indices limits achievable optical efficiency. We address this challenge by designing a multi-level photonic unit cell with a metallic overlayer and optimizing it for polarization-independent beam steering

(Figure 5a). The unit cell design consists of an array of polymeric posts with identical widths but varying heights, over which a metallic layer is deposited. Considering the directional nature of material deposition, the geometry was tailored such that the metallic overlayer slightly extends over the sidewalls of the posts as well as their top surfaces. Next, we formulate a multi-objective photonic design optimization aimed at maximizing steering efficiencies under both TE and TM polarizations. This dual-polarization optimization is key, as the orientation of the photonic surface dynamically changes due to the rotation of the magneto-photonic metaparticle, resulting in a superposition of TE and TM components in the incident light. We employ a genetic algorithm (GA) to perform the optimization with rigorous coupled-wave analysis (RCWA) used to calculate steering efficiency. The GA evolves a population of candidate structures by evaluating their fitness with respect to both objectives, selecting high-performing designs, and generating new candidates through crossover and mutation (Figure 5b). This process continues until convergence toward a Pareto-optimal set of solutions where improvements in one objective cannot be achieved without compromising the other. Among the final candidates, the design with the maximum product of efficiencies under TE and TM polarizations was selected.

For our design, the parameters were set for an operating wavelength of 633 nm, a steering angle of 50°, and ten posts per unit cell whose heights served as the optimization variables (Figure 5a). This optimization method generates a set of solutions at each generation and progressively evolves the Pareto front, driving the steering efficiency toward unity for both TE and TM modes (Figure 5c). From the Pareto front at each generation, the knee point is selected as the optimal solution by minimizing the distance to the ideal point (TE, TM) = (1, 1). This optimal solution steadily increased and eventually converged over successive generations (Figure 5d). Because the optimization simultaneously targets high steering efficiency for both TE and TM polarizations, the evolution of their Pareto front reflects the trade-off surface between the two objectives. The evolution of the Pareto front and the growing number of optimal (rank-1) candidates further confirm the effective progression of the multi-objective optimization (Figure S4). The resulting optimized structure in air achieves steering efficiencies of 0.77 and 0.79 for TE and TM modes, respectively (Figure 5e). These efficiencies are significantly higher than for conventional gratings: for comparison, a conventional ruled grating with a blazing angle of 25° yields a lower efficiency of 0.70 for TM mode, and a much lower efficiency of ~0.50 for TE mode. This highlights the superior polarization-independent performance of the optimized structure enabled by our design and fabrication strategy. Furthermore, the optimized unit cell design retains high steering efficiencies of 0.76 (TE) and 0.79 (TM) even in an aqueous environment ($n = 1.33$) (Figure 5f and Figure S5). This highlights the versatility of our magneto-photonic design for high-performance beam steering in different environments.

### 3. Discussion
We have introduced a magneto-photonic metaparticle as a versatile and mobile platform for real-time beam steering. By integrating magnetic actuation with nanoimprinted photonic surfaces, the metaparticle platform overcomes the limitations of static metasurfaces and the limited optical functionality of conventional particle scatterers, achieving precise locomotion and orientation control in an aqueous environment. Experimental results demonstrate the feasibility of programmable motion, decoupled rotational and translational control, and dynamic light steering in direct response to external magnetic fields. The synchronized magnetic and photonic orientations enable the metaparticle to function as a reconfigurable micro-optical element capable of real-time, directional light control.

We envision several promising directions stemming from this work. The large degree of freedom in designing photonic surfaces can enable a wide range of photonic functionalities. While this study focuses on creating a photonic surface for beam steering, these functionalities can be extended to beam focusing or diverging for microlens applications, as well as to spectral control, polarization control, and emission control for micro-filter applications. Moreover, imprinting not only ensures the scalability of nanoscale patterning but also allows for three-dimensional structures to be fabricated, as demonstrated by the structures used in this study. The broad design space could be further expanded by incorporating sharp or curved complex surfaces, and varying the height of metastructures could be fabricated by grayscale lithography[38]. The metaparticle design is inherently scalable, and with appropriate control of the photonic pattern size, the structure could be miniaturized to tens or even several micrometers, thereby extending its applicability to diverse media, including biological and chemical environments. This miniaturized metaparticle platform holds strong potential for biophotonic applications. Operating at wavelengths within the biologically transparent windows can enable deep tissue interaction with minimal scattering and absorption. The first near-infrared (NIR-I) window (650–950 nm) offers reduced light scattering compared to the visible range, and is widely used for high-resolution fluorescence imaging, photothermal therapy, and optogenetic stimulation at shallow to moderate tissue depths.[17,39,40] The second near-infrared (NIR-II) window (1000–1350 nm) provides even lower scattering and absorption, allowing light to penetrate deeper into biological tissue while maintaining high signal-to-background ratios, which is beneficial for deep-tissue imaging, targeted photothermal heating, and photodynamic therapy.[41,42] By tuning the magneto-photonic metaparticles to operate in these spectral regions, it may be possible to directly deploy them in biological tissues for targeted, minimally invasive therapeutic interventions or real-time in vivo optical sensing.

## 4. Experimental Section

**4-1 Fabrication of magnetic microdiscs**. Diluted solution of negative photoresist (SU-8 2010, MicroChem) is spincoated onto a silicon wafer to form 800-nm-thickness SU-8 bottom layer. A photomask which consists of square arrays of 100-μm-diameter circles was used to regioselectively irradiate the SU-8 film with the mask aligner (MABA6, SUSS). SU-8 developer is used to create SU-8 microdiscs as a bottom layer. Cobalt cores with 50 nm thickness are then deposited onto SU-8 microdiscs using the lift-off process. SU-8 is spincoated and developed onto a Co-deposited SU-8 microdiscs to create a Co-incorporated SU-8 microdisc. The samples were then immersed inside the MF-319 solution to detach microdiscs from a substrate.

**4-2 Fabrication of magneto-photonic microdiscs**. Before detaching magnetic microdiscs, surfaces of SU-8 microdiscs are imprinted with reflective ruled gratings (300 grooves/mm or 1200 grooves/mm, Thorlabs) at 90 °C and 150 psi for 300s (NX B200, Nanonex). 50 nm of Ag is deposited using e-beam evaporator (CHA Industries). The samples were then immersed inside the MF-319 solution to detach microdiscs from a substrate.

**4-3 Magnetic actuation procedure of magnetic particle rotation.** To facilitate magnetic control of rotation of a magnetic microdisc, a neodymium magnet (max energy product: 263-287 kJ/m$^3$) is precisely manipulated to induce in-plane and out-of-plane rotation motion of the microdisc. In-plane rotation is actuated by moving the magnet in a circular trajectory around the center of the microdisc while gradually rotating the magnet to make magnetic field continuously face the microdisc. However, when the microdisc is in contact with the substrate, frictional forces can be high enough to prevent in-plane rotation, even when appropriate magnetic actuation is applied. In such cases, out-of-plane rotation can serve to monetarily lift the disc off the substrate, reducing contact and enabling smoother in-plane rotation. Out-of-plane rotation is actuated by fixing the position of the magnet and rotating it around its central axis, which is parallel to the substrate surface.

**4-4 Characterization**. Images of microdiscs and rotational motions of magnetic microdiscs are observed with optical microscopy in a reflection mode, where a 5× and 10× objective lens with a numerical aperture of 0.30 and 0.45, respectively, were used. Magnetic moments of magnetic cores at different orientations are characterized with vibrating-sample magnetometer.

**4-5 Simulation and optimization**. Calculations on diffraction efficiencies and diffraction orders were performed with the RCWA method, using an open-source software package[43]. Multi-objective optimization was performed with Global Optimization Toolbox implemented in MATLAB.

**Data Availability Statement**
The data that support the findings of this study are available from the corresponding author upon reasonable request.

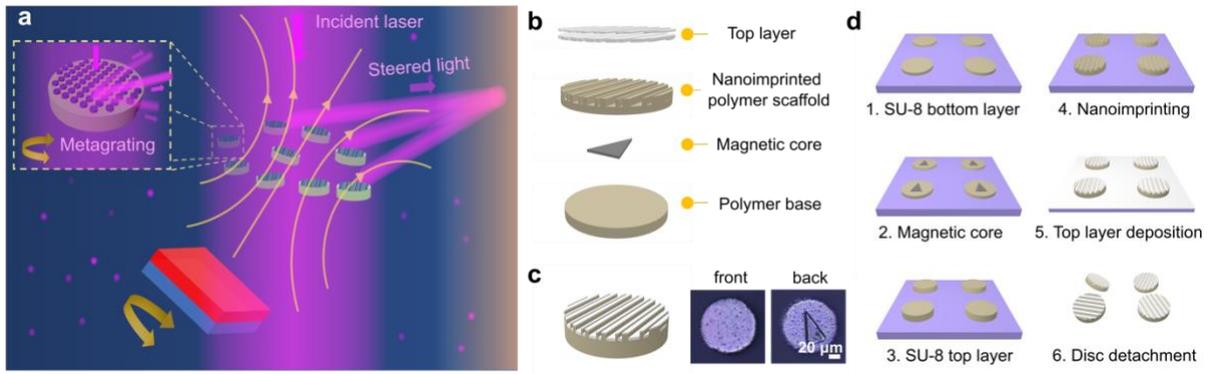

**Figure 1.** Concept, structure, and fabrication of magneto-photonic metaparticles. (a) Schematic illustration showing a metaparticle with a photonic surface pattern and an embedded magnetic core. This hybrid design merges the reconfigurability of photonic metasurfaces with the mobility of magnetic actuation, enabling programmable collective motion and beam steering. (b) Structural anatomy, (c) illustration and optical microscope image, and (d) fabrication process of an integrated magneto-photonic metaparticle.

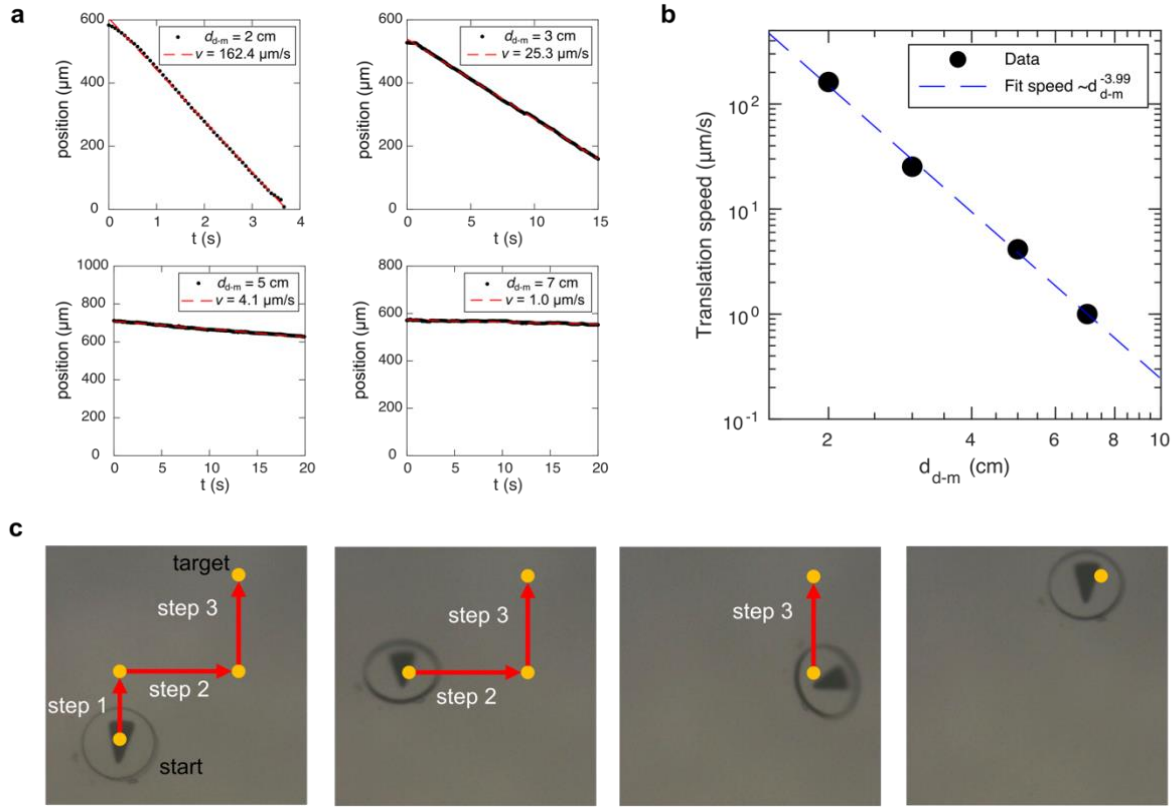

**Figure 2.** Controlled locomotion of magnetic metaparticles. (a) Measured velocities of magnetic microdiscs as a function of distance ($d_{d\text{-}m}$) from the magnet. (b) Log-log plot of measured velocity ($v$) as a function of distance ($d_{d\text{-}m}$). The trend is consistent with the $v \propto d_{d\text{-}m}^{-4}$ scaling, predicted from the far-field magnetic dipole gradient ($dB/d(d_{d\text{-}m}) \propto d_{d\text{-}m}^{-4}$) and viscous drag balance for a saturated ferromagnetic inclusion in low-Reynolds-number flow. (c) Sequential locomotion of a magnetic particle programmed to follow a series of translational motions (see also Supplementary Movie S1).

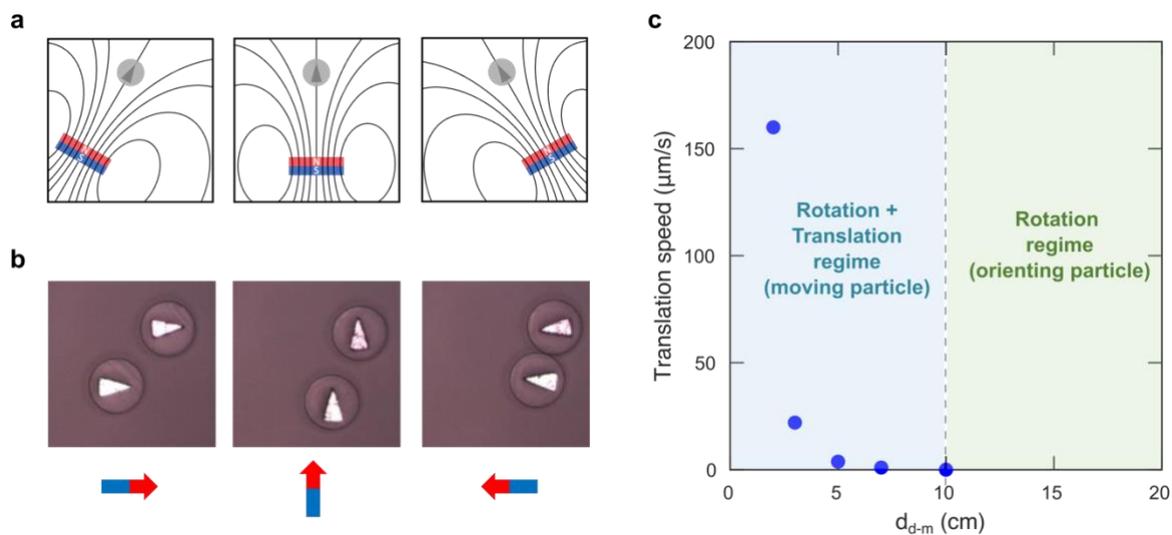

**Figure 3.** Rotational motion and field-dependent motion regimes of magnetic metaparticles. (a) Illustration of particle alignment under different magnet orientations and positions. (b) Actual alignment of particles under different directions of the applied magnetic field (arrows indicate field direction). (c) Distinct motion regimes are controlled by the field strength: moving (high-field / closer proximity to the magnet) vs orienting (low-field / larger distance to the magnet). (See also Supplementary Movie S2).

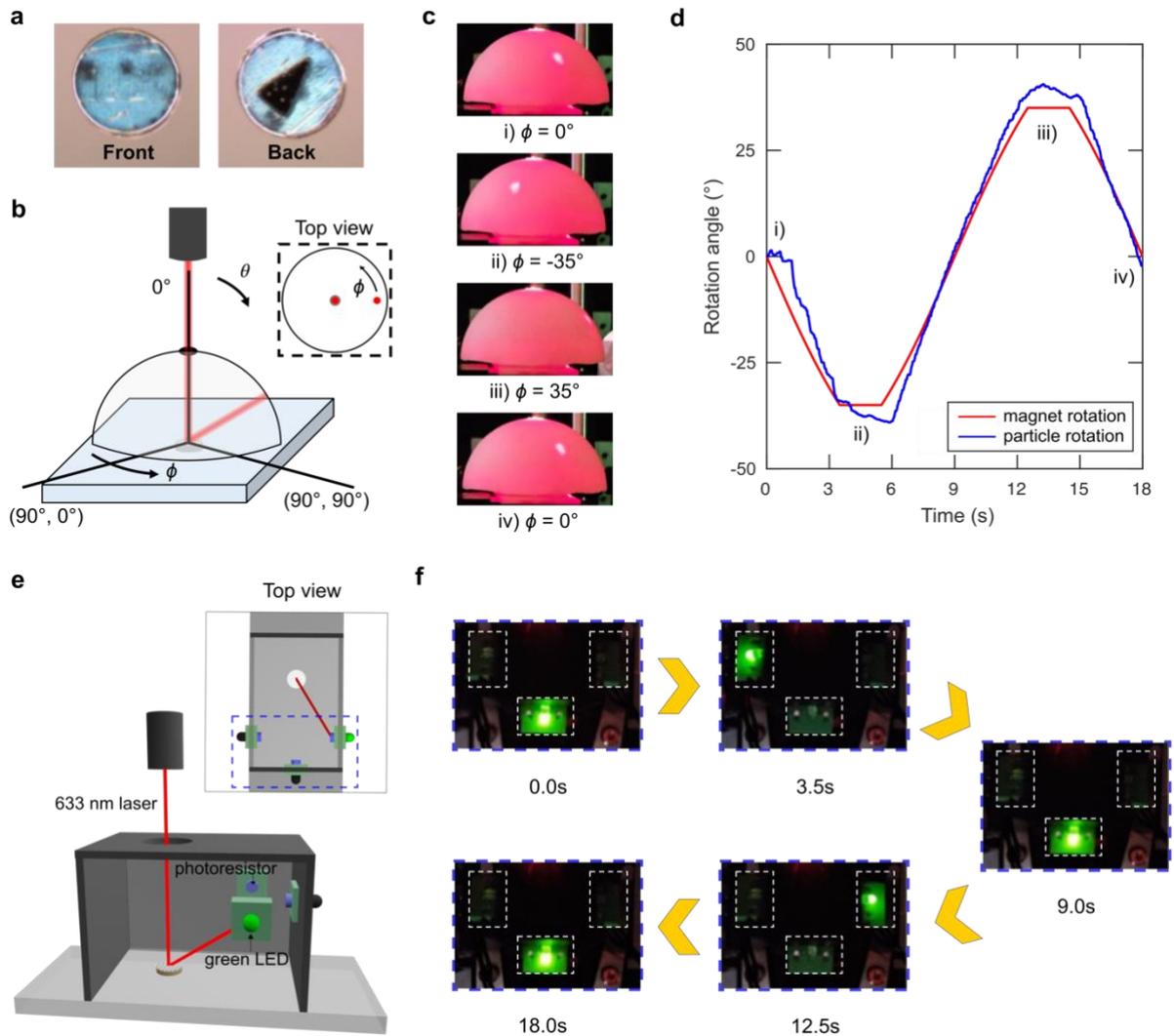

**Figure 4.** Integration of optical functionality and demonstration of active beam steering. (a) OM images of magneto-photonic microdiscs with a nanoimprinted grating surface. (b) Experimental setup for characterizing active light-steering behavior under controlled circular magnet motion. (c) Trajectories of the steered beam as the microdisc is rotated by 35° clockwise, 70° counterclockwise, and again 35° clockwise. (d) Synchronized response of particle rotation and magnet rotation (Supplementary Movie S4). (e) Setup for real-time beam-steering demonstration: a container with edge light sensors (inward-facing photodetectors and outward-facing indicator LEDs). When the steered beam from the magneto-photonic particle strikes a light sensor, the corresponding LED is activated. (f) Top view of the setup in (e), showing the progression of beam steering under programmed magnet rotation of a particle immersed in water. (See also Supplementary Movie S5).

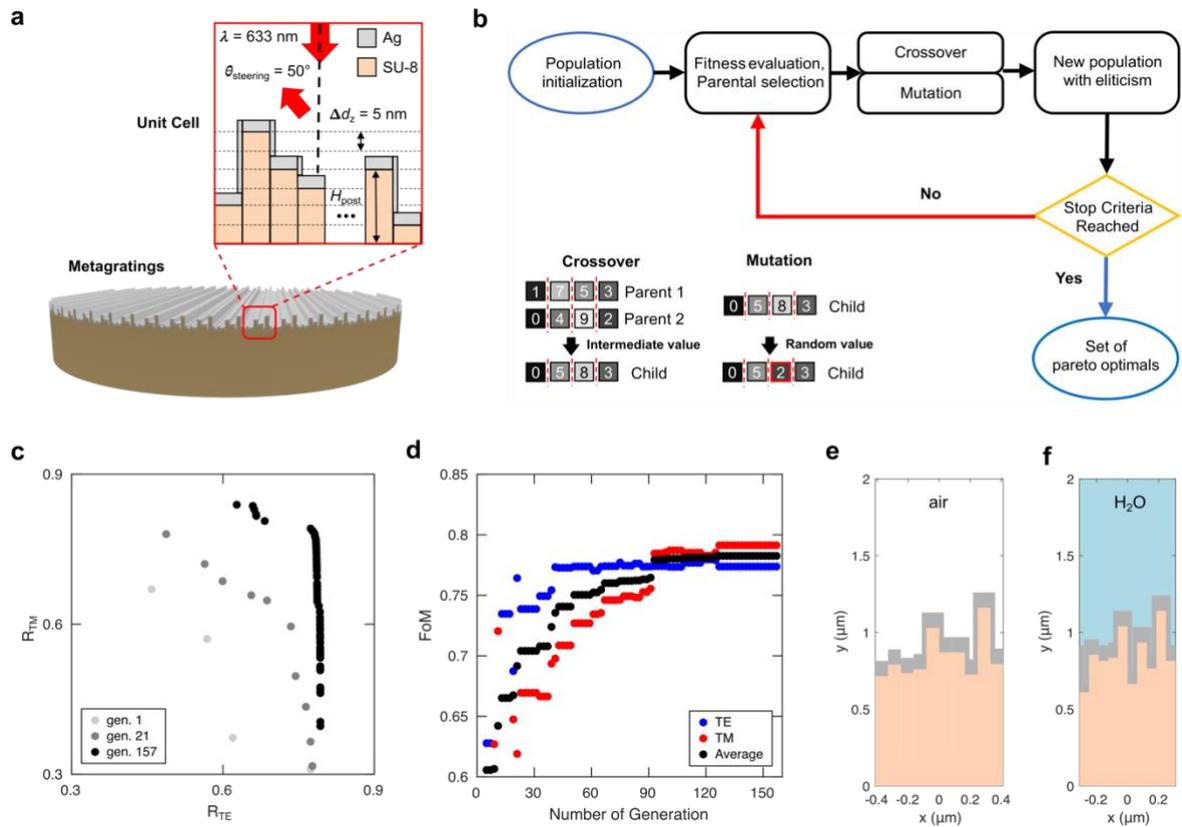

**Figure 5.** Nanoimprint-compatible design of metaparticles for polarization-insensitive light steering. (a) Illustration of a metagrating-engraved metaparticle and unit cell structure consisting of polymeric (SU-8) posts with a metal (Ag) overlayer. Wavelength is 633 nm and target steering angle is 50°. (b) Flow chart of a genetic-algorithm-implemented multi-objective optimization. (c) Evolution of the Pareto front at the 1st, 21st, and 157th generations of the genetic multi-objective optimization. (d) Multi-objective optimization over successive generations, with the Figure of merit defined by the knee point that minimizes the distance to the ideal point (TE, TM) = (1, 1). The average steering efficiency shows a gradual increase over generations. (e) Optimized unit cell in air (n = 1.0) achieving steering efficiencies of 0.77 (TE) and 0.79 (TM). (f) Optimized unit cell in water (n = 1.33), achieving efficiencies of 0.76 (TE) and 0.79 (TM).